\documentstyle[epsfig,12pt]{aipproc}

\input epsf

\begin{document}
\title{The STARE Project: \\A Progress Report}

\author{C.A.~Katz$^*$, J.N.~Hewitt$^*$, C.B.~Moore$^*$, and
B.E.~Corey$^\dagger$} 
\address{$^*$MIT Department of Physics and Research Laboratory of
Electronics \\ Cambridge, MA 02139 \\
$^\dagger$Haystack Observatory, Off Route 40, Westford, MA 01866}

\maketitle

\begin{abstract}

  The Survey for Transient Astronomical Radio Emission (STARE) is a
  wide-field monitor for transient radio emission at 611~MHz on
  timescales of fractions of a second to minutes.  Consisting of
  multiple geographically separated total-power radiometers which
  measure the sky power every 0.125~sec, STARE has been in operation
  since March~1996.  In its first seventeen months of operation, STARE
  collected data before, during, and after 173 gamma-ray bursts.  Seven
  candidate astronomical radio bursts were detected within $\pm 1\,\rm
  hr$ of a GRB, consistent with the rate of chance coincidences expected
  from the local radio interference rates.  The STARE data are therefore
  consistent with an absence of radio counterparts appearing within $\pm
  1\,\rm hr$ of GRBs, with $5\sigma$ detection limits ranging from tens
  to hundreds of kJy.  The strengths of STARE relative to other radio
  counterpart detection efforts are its large solid-angle and temporal
  coverage.  These result in a large number of GRBs occuring in the
  STARE field of view, allowing studies that are statistical in nature.
  Such a broad approach may also be valuable if the GRBs are due to a
  heterogenous set of sources.

\end{abstract}

\section*{Introduction}

  The Survey for Transient Astronomical Radio Emission (STARE) is a
  project designed to detect transient radio signals at 611~MHz on
  timescales of fractions of a second to minutes.  Local interference
  rejection is accomplished by using geographically separated multiple
  detectors and a coincidence requirement.  STARE monitors a large solid
  angle 24~hours/day, with an operating efficiency of $\sim\!95\%$.
  We present here a brief description of the project and some of
  the results from the first 17~months of operation.

\section*{Instrumentation}

\begin{figure}
\begin{center}
\epsfysize=0.75\textheight
\epsfbox{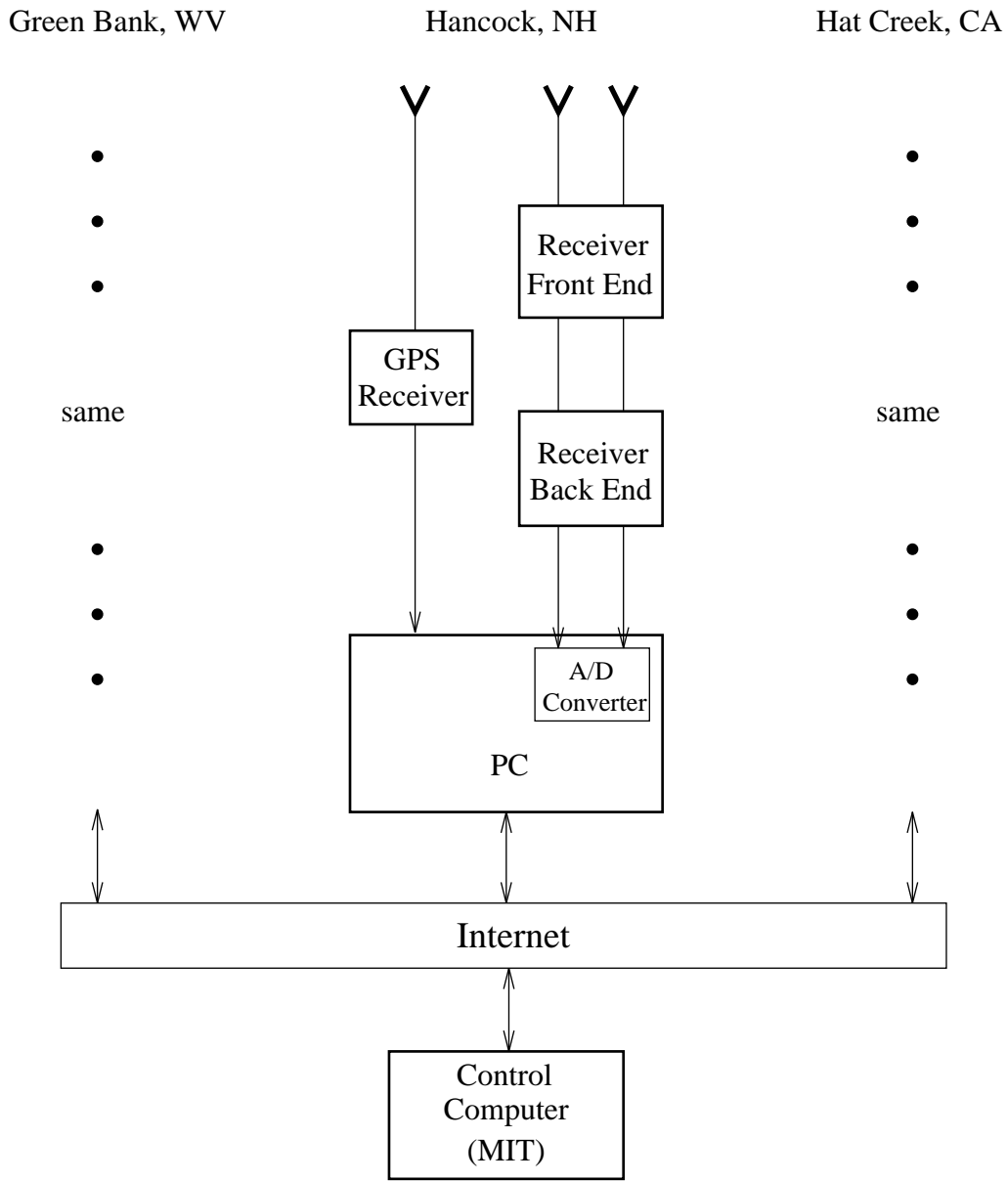}
\bigskip
\caption{STARE System Organization\label{F:ckatz:1}}
\end{center}
\end{figure}

  STARE consists of three detectors located at geographically separated
  sites: Hancock, NH, Green Bank, WV, and Hat Creek, CA\@.  At each site
  is a total power radiometer, a GPS receiver which provides timing
  information, and a PC which controls the equipment.  Operation is
  fully automated, and is coordinated over the internet by a computer at
  MIT, which also receives data from the sites.  The organization of the
  system is shown in Figure~\ref{F:ckatz:1}.

  The radiometers are simple single-conversion,
	dual-polarization receivers.  System specifications:
	\begin{itemize}
  	  \item Center Frequency: 611 MHz
	  \item Bandwidth       :   4 MHz
    	  \item Beam Solid Angle: 1.5 sr
    	  \item Integration Time: $20\mu \rm s$ to $0.125\,\rm s$
    	  \item RMS Sensitivity : $\sim\!3\,\rm kJy$ (zenith,
	                          $0.125\,\rm s$ averaging)
	\end{itemize}

\section*{Transient Detection by Coincidence Requirement}

  Radiofrequency interference (RFI) presents a major obstacle
  to transient detection at UHF\@.  At these frequencies it is
  imperative that an experiment have an interference rejection scheme.
  STARE filters out RFI using a coincidence requirement.  To be
  identified as astronomical, a signal must appear in at least two of
  the detectors simultaneously.  The power of this criterion is apparent
  from the STARE results: in one year of operation, the Hancock station
  recorded 78,714 individual events (instances in which the radiometer
  output exceeded the baseline by at least $5\sigma$) while the
  Green~Bank station recorded 260,407.  Out of these, only 138
  coincidences were identified, yielding a rejection rate of well over
  99\%.

  With such a scheme, there is always the possibility of
  coincidences due to chance.  Using the mean event rates measured for
  Hancock ($7\,{\rm hr}^{-1}$) and Green~Bank ($13\,{\rm hr}^{-1}$), the
  mean time between chance coincidences is found to be about 4~days,
  though the chance coincidence rate is quite variable since it depends on
  the (variable) underlying rates of RFI bursts at each site.
  Adding a third site with a rate identical to that of Green Bank
  increases the mean time between chance coincidences to about 27~years.
  For this reason STARE was designed to include three sites.  The
  Hat~Creek site, however, has been found to have an RFI environment
  unsuitable for general transient detection.  Work is in progress to
  remedy this situation in order to give STARE the full interference rejection
  capability for which it was designed.

\newpage

\section*{Survey for Radio Counterparts to Gamma-Ray Bursts}

  A strength of the STARE project in searching for GRB
  counterparts is that it monitors a large fraction of the sky
  nearly twenty-four hours per day.  It will record from GRBs in its
  field of view any radio emission occurring after, during, or even
  before the GRB itself (with intensity above the STARE sensitivity
  level, of course).  The price paid for a large field of view is lower
  sensitivity than that of a narrow-field detector.  However, the large
  field of view also increases the number of GRBs expected to occur
  within the antenna beam.

\begin{figure}
\begin{center}
\epsfysize=0.32\textheight
\epsfbox{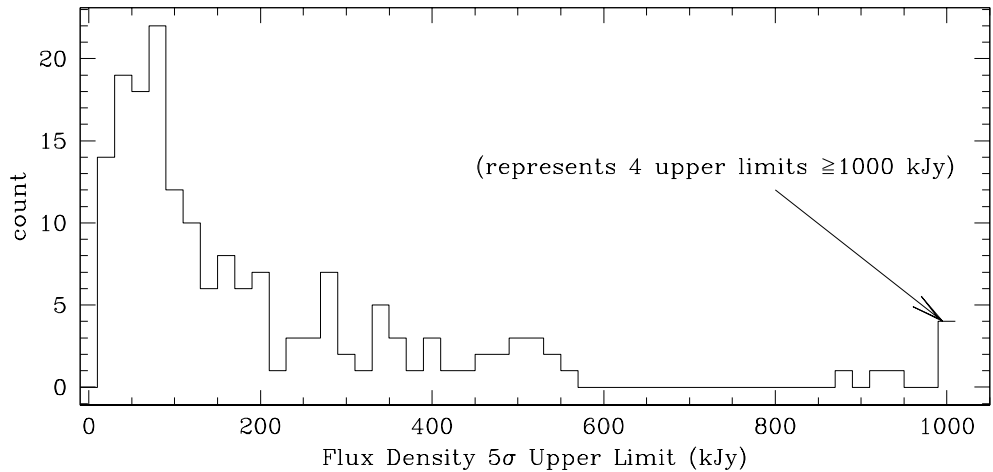}
\bigskip
\caption{Histogram of STARE flux density $5\sigma$ upper limits for 173
GRBs\label{F:ckatz:2}}
\end{center} 
\end{figure}

  STARE began multi-site operation on 26~March~1996 when the Green~Bank
  station came on-line (Hancock came on-line in August~1995).  Between
  that time and 31~August~1997, 173~GRBs (detected by BATSE) occurred
  in the field of view (above $20^\circ$ elevation) of at least one of
  the STARE sites.  For each GRB, the STARE coincidence record was
  examined for detections within $\pm 1\,\rm hr$ of the gamma-ray event.
  7 positive results were found, which is consistent with the number
  expected from chance.  None of the seven occurred simultaneously with
  a GRB\@.  In addition, for each GRB, the raw STARE data record from each
  site was examined manually for unusual activity within $\pm 30\,\rm
  min$ of the GRB\@.  Nothing was found which was not obviously the
  usual RFI\@.  We conclude that {\bf the STARE data are consistent with
  an absence of radio counterparts appearing within $\mathbf{\pm 1\,
  hr}$ of GRBs, with $\mathbf{5\sigma}$ detection limits ranging from
  tens to hundreds of kJy.}  A histogram of the ensemble of upper limits
  is shown in Figure~\ref{F:ckatz:2}.

\end{document}